\RequirePackage{fix-cm}
\documentclass[smallcondensed,natbib]{svjour3}     % onecolumn (ditto)
\smartqed  % flush right qed marks, e.g. at end of proof
\usepackage{graphicx, xcolor}
\usepackage{subfig}
\usepackage{amsmath,amssymb}
\usepackage[pdftex,breaklinks,colorlinks,urlcolor=cyan,citecolor=blue,linkcolor=blue]{hyperref}
 \journalname{Space Sci Rev}
\begin{document}

\title{$^3$He-rich solar energetic particles: Solar sources%\thanks{Grants or other notes
%about the article that should go on the front page should be
%placed here. General acknowledgments should be placed at the end of the article.}
}
%\subtitle{Do you have a subtitle?\\ If so, write it here}

%\titlerunning{Short form of title}        % if too long for running head
%\dedication{in memory of Professor Karel Kudela}

\author{Radoslav Bu\v{c}\'ik%         \and
       % Second Author %etc.
}

%\authorrunning{Short form of author list} % if too long for running head

\institute{R. Bu\v{c}\'ik \at 
            Southwest Research Institute, San Antonio, TX 78238, USA \\
            \email{radoslav.bucik@swri.org}  \and \at
             Max-Planck-Instit\"ut f\"ur Sonnensystemforschung, D-37077, G\"ottingen, Germany
             \and \at Instit\"ut f\"ur Experimentelle and Angewandte Physik, Christian-Albrechts-Universit\"at zu Kiel, D-24188, Kiel, Germany %\\
%              Tel.: +123-45-678910\\
%              Fax: +123-45-678910\\
            %  \email{radoslav.bucik@swri.org}           %  \\
            % \emph{Present address:} of F. Author  %  if needed
%           \and
%          S. Author \at
%              second address
}

\date{Received: 3 January 2020 / Accepted: 21 February 2020}
% The correct dates will be entered by the editor

\maketitle

\begin{abstract}
$^3$He-rich solar energetic particles (SEPs), showing up to a 10,000-fold abundance enhancement of rare elements like $^3$He
or ultra-heavy nuclei, have been a puzzle for more than 50 years. One reason for the current lack of understanding of 
$^3$He-rich SEPs is the difficulty resolving the source regions of these commonly occurring events. Since their discovery, there has been strong evidence
that $^3$He-rich SEP production is associated with flares on the Sun. Anomalous abundances of $^3$He-rich SEPs have been 
attributed to a unique acceleration mechanism that must routinely operate at flare sites. Flares associated with $^3$He-rich 
SEPs have been often observed in jet-like forms indicating an acceleration in magnetic reconnection involving field lines open
to interplanetary space. Owing to a fleet of spacecraft around the Sun, providing a greatly improved resolution of solar imaging
observations, $^3$He-rich SEP sources are now explored in unprecedented detail. This paper outlines the current 
understanding of $^3$He-rich SEPs, mainly focusing on their solar sources.

 %\PACS{96.40.Fg \and PACS code2 \and more}
% \subclass{MSC code1 \and MSC code2 \and more}
\end{abstract}

\section{Introduction}
\label{intro}
$^3$He-rich solar energetic particles (SEPs) were discovered in the early 1960s \citep{1962PhRvL...8..389S} and since then they have been the focus 
of many experimental \citep[e.g.,][]{1970ApJ...162L.191H} and theoretical investigations \citep[e.g.,][]{1978ApJ...224.1048F}. As their 
name indicates, $^3$He-rich SEPs are characterized by the enormous enrichment of the rare isotope $^3$He by factors of up to 10$^4$ above the solar wind or coronal abundance \citep[$^3$He/$^4$He$\sim$4$\times$10$^{-4}$ in the solar wind;][]{1998SSRv...84..275G}. Heavy ($^{22}$Ne -- $^{56}$Fe) and ultra-heavy ions (mass $>$70\,amu) are enhanced by a factor of 3--10 and $>$100, respectively, independently of the amount of $^3$He enhancement \citep{1986ApJ...303..849M,1994ApJS...90..649R}. It has been interpreted as evidence that different mechanisms are involved in the acceleration of the $^3$He and the heavy ions. The abundance enhancement factor increases with atomic number (or mass) for heavier ions approximately as a power-law \citep{2004ApJ...606..555M,2004ApJ...610..510R}. Furthermore, whereas with He the lighter isotope is enhanced, with heavier ions (e.g., Ne, Mg) heavier isotopes are enhanced \citep[see][for a review]{1984SSRv...38...89K,2007SSRv..130..231M}. $^3$He-rich SEP events are a ubiquitous phenomenon with the rate of occurrence corresponding to $\sim$10$^3$ events/yr ($\sim$1\,MeV\,nucleon$^{-1}$ $^3$He/$^4$He$>$0.1) on the visible solar disk at solar maximum \citep{1994ApJS...90..649R,2012ApJ...759...69W}. 

Systematic research of $^3$He-rich SEP sources started more than a decade ago using the {\sl Solar Heliospheric Observatory} (SOHO) extreme ultraviolet (EUV) imaging observations \citep{2006ApJ...639..495W,2006ApJ...650..438N,2006ApJ...648.1247P}. Recent reviews on $^3$He-rich SEP events \citep{2007SSRv..130..231M,2017LNP...932.....R} focus more on energetic ion characteristics obtained from in-situ measurements. In this paper we present new insights on $^3$He-rich SEP sources provided by recent space missions such as the {\sl Solar TErrestrial RElations Observatory} with two progressively separating spacecraft STEREO-A, -B, and the {\sl Solar Dynamics Observatory} (SDO) with unprecedentedly high-resolution observations.

\section{Soft X-rays flares}
\label{sec:1}
$^3$He-rich SEP events have been associated with minor (low intensity) soft X-ray (1--8\,{\AA}) GOES (mostly B- and C-class) flares \citep[e.g.,][]{2006ApJ...650..438N,2015ApJ...806..235N,2009ApJ...700L..56M,2016ApJ...833...63B}. Also an inverse correlation between the soft X-ray peak intensity and $^3$He (ultra-heavy) enrichment has been reported \citep{1988ApJ...327..998R,2004ApJ...610..510R}. This has been explained by arguing that in small flares the limited available energy is almost all absorbed by the rare $^3$He and the heavy ions, while in large flares there is enough energy to accelerate more abundant elements which decreases the relative enrichment of the rare species \citep{2004ApJ...610..510R}. $^3$He-rich events have been termed impulsive SEP events based on the time-intensity profile of the associated X-ray flares. The $^3$He-rich SEPs and X-rays may originate in different regions in the corona because ions are accelerated at reconnection sites on open field lines while flares involve reconnection on neighboring closed loops \citep{2014SoPh..289.3817R}. Often $^3$He-rich SEP events are found without an observed X-ray flare \citep{1987SoPh..107..385K}, showing a signal only in EUV ($\sim$100--200\,{\AA}) wavelengths \citep[e.g.,][]{2006ApJ...650..438N}. 

\begin{figure*}[t!]
\centering
$\begin{array}{rl}
    \includegraphics[trim=0 20.5 0
0,clip=true,width=0.6\textwidth]{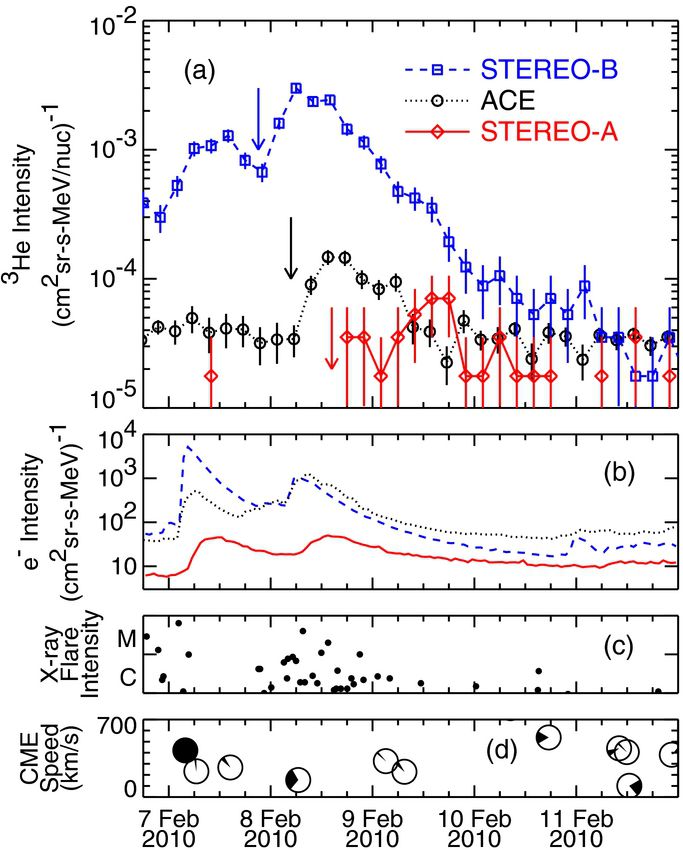} &
    \includegraphics[trim=0 7.3 41.5
5.5,clip=true,width=0.36\textwidth]{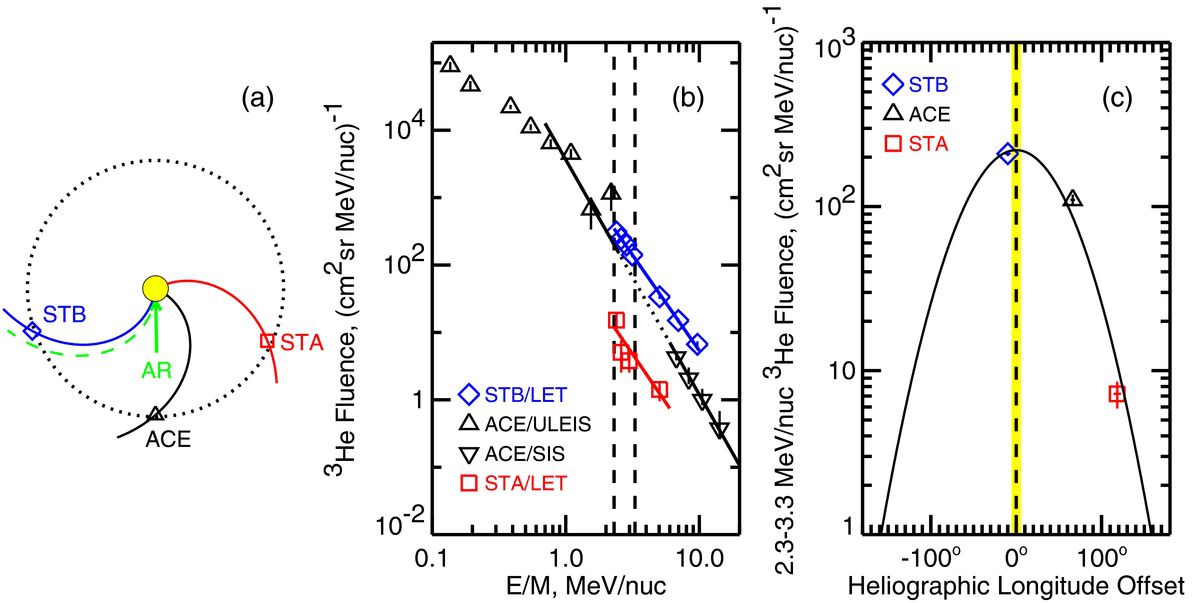}\\
    \multicolumn{2}{l}{\includegraphics[trim=0 0 0
38,clip=true,width=0.6\textwidth]{apj452161f2_hr.jpg}}
\end{array}$
\caption{(Left) The 2010 February 7 $^3$He-rich SEP event measured on angularly separated spacecraft STEREO-A, -B ($\sim$3\,MeV\,nucleon$^{-1}$) and ACE ($\sim$9\,MeV\,nucleon$^{-1}$). (Right) The location of spacecraft and the source active region. Adapted from \citet{2013ApJ...762...54W}.}
\label{fig:1} 
\end{figure*}

\section{Cone of ions emission} 
\label{sec:2}
Flares associated with $^3$He-rich SEP events observed at the Earth were found to come from a limited region ($\sigma$$\sim$16$^{\circ}$) on the western hemisphere ($\sim$W57) of the Sun which is magnetically well-connected to the observer \citep[e.g.,][]{1999SSRv...90..413R}. Imaging observations on the spacecraft (STEREO-A) angularly separated from Earth (ACE) have shown a significantly broader longitudinal distribution of source flares \citep[$\sigma$$\sim$42$^{\circ}$;][]{2016ApJ...833...63B}, inconsistent with simple interplanetary magnetic field spiral approximation even when combining with divergent coronal field lines from potential-field source-surface model \citep{2015ApJ...806..235N}. The sensitivity of ion telescopes can affect the longitudinal distribution - the weak ion signal measured by ACE would have been below the threshold of past instruments. A wide cone of emission can be expected from measurements of $^3$He-rich SEP events on angularly separated spacecraft \citep{2013ApJ...762...54W}. Figure \ref{fig:1} shows the 2010 February 7 $^3$He-rich SEP event measured at three widely ($>$60$^{\circ}$) separated spacecraft STEREO-A, -B, and ACE.

Several mechanisms have been discussed that contribute to the measured broad spatial distribution of $^3$He-rich SEPs including large-scale coronal waves (c.f. Section \ref{sec:4}), field lines meandering due to supergranular motions in the photosphere \citep[e.g.,][]{2012ApJ...751L..33G,2017ApJ...846..107Z} where a small percentage of the field lines can make quite large excursions \citep[e.g.,][]{1969ApJ...155..777J}, or separatrix S-web \citep{2018ApJ...869...60S} where field lines from a narrow corridor, connecting low-latitude coronal holes with polar coronal holes, map to 
the extended longitudes in the heliosphere \citep[e.g.,][]{2017ApJ...840L..10H}. 

\begin{figure*}[t!]
\centering
$\begin{array}{rl}
    \includegraphics[trim=0 0 130
0,clip=true,width=0.51\textwidth]{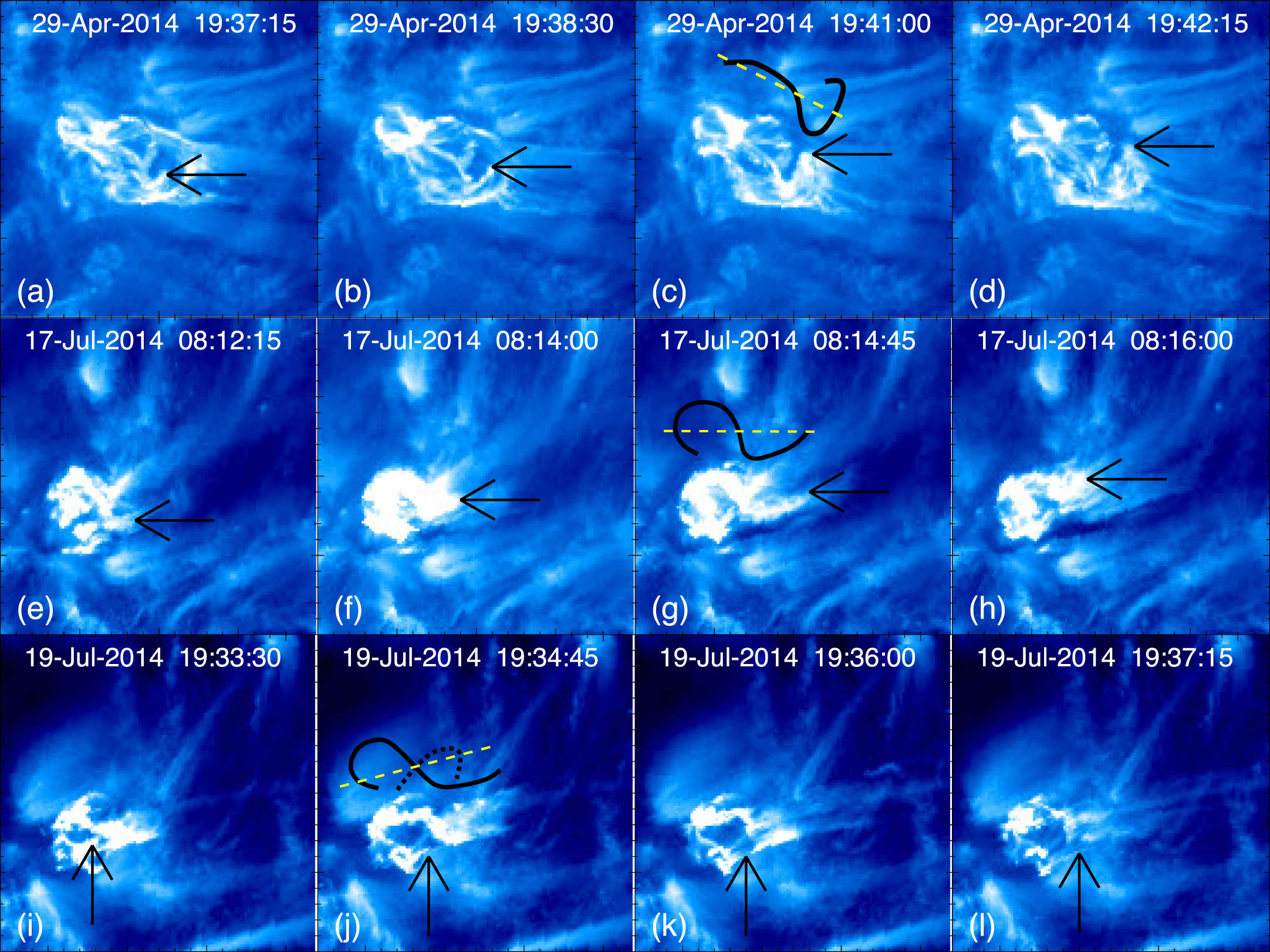} &
    \includegraphics[trim=240 0 0
260,clip=true,width=0.47\textwidth]{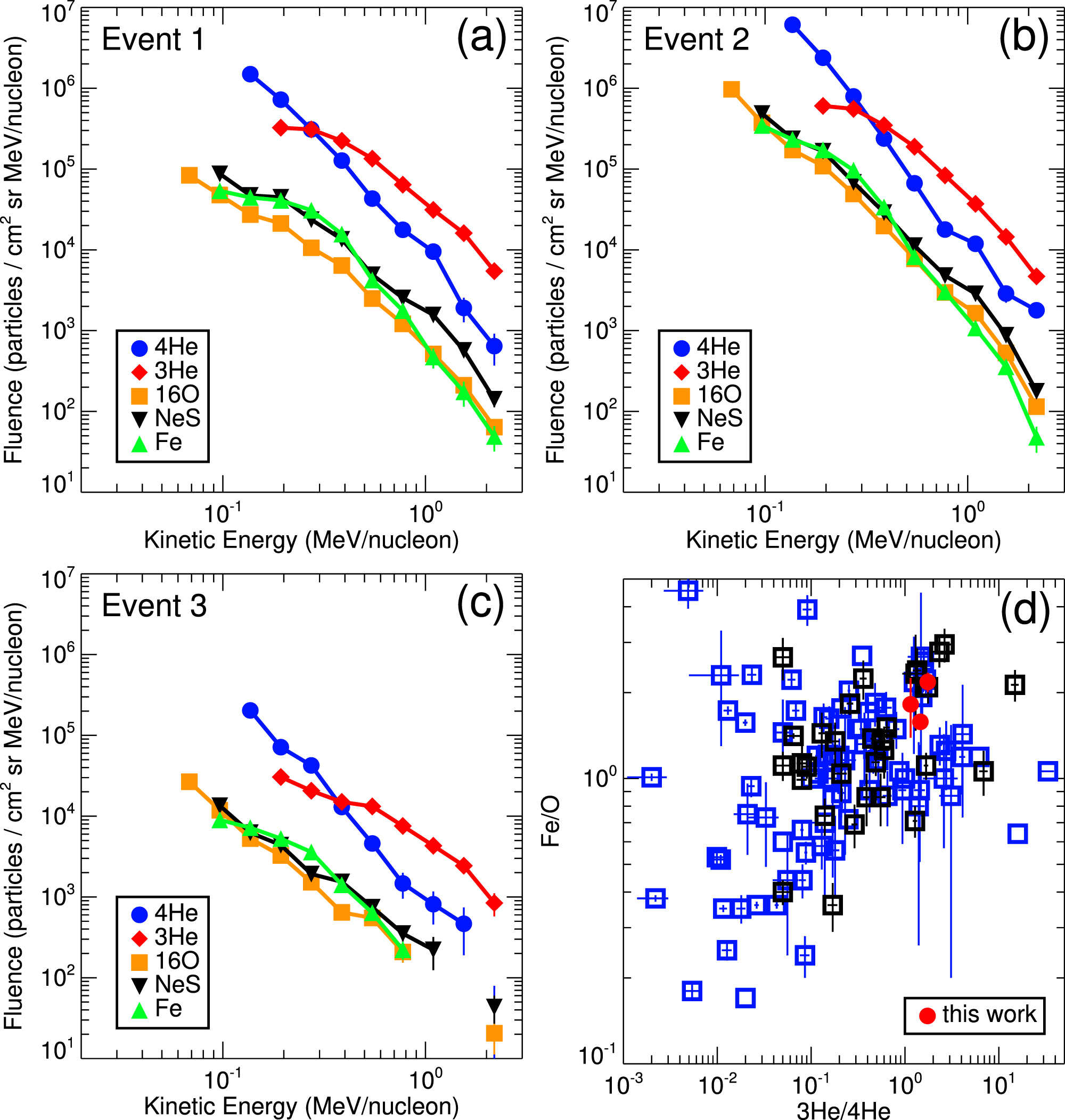}
\end{array}$
\caption{(Left) STEREO-A EUV 171\,{\AA} images of the helical jet evolution in $^3$He-rich SEP events on 2014 April 29, 2014 July 17, and 2014 July 19. (Right) Fe/O vs. $^3$He/$^4$He for events in the left panel (red circles) and all (109) previously reported $^3$He-rich SEP events (blue and black squares) at the energy $\sim$0.1--2\,MeV\,nucleon$^{-1}$. Adapted from \citet{2018ApJ...852...76B}.}
\label{fig:3} 
\end{figure*}

\section{Jets} 
\label{sec:3}
Early studies of $^3$He-rich SEP events showed an association with small area H$\alpha$ (6563\,{\AA}) 'visible' flares \citep{1978ApJ...225..281Z,1985ApJ...292..716R,1987SoPh..107..385K}. Further studies have shown that flares associated with $^3$He-rich SEPs are commonly observed as collimated or jet-like forms in EUV/X-ray images \citep{2006ApJ...650..438N,2008ApJ...675L.125N,2015ApJ...806..235N,2014ApJ...786...71B,2018ApJ...852...76B,2015A&A...580A..16C,2020ApJS..246...42W}, with a high-altitude extension in white-light coronagraphs as narrow coronal mass ejections (CMEs) \citep{2001ApJ...562..558K,2006ApJ...639..495W,2014SoPh..289.4675R,2012ApJ...759...69W,2016A&A...585A.119W,2018A&A...619A..34B}. The events with higher $^3$He enrichment are coupled with narrow CMEs with low speeds much more strongly than events with lower enrichment \citep{2014SoPh..289.4675R}. Several events, however, show only an amorphous EUV brightening without a jet that is surmised to be an instrument resolution issue \citep[e.g.,][]{2006ApJ...639..495W} and perhaps a projection effect. Jets result from magnetic reconnection between field lines open to interplanetary space and emerging magnetic flux \citep{1992PASJ...44L.173S} or erupting mini-filament \citep{1996ApJ...464.1016C,2015Natur.523..437S}. Thus, an association of $^3$He-rich SEPs with jets implies an ion acceleration mechanism related to magnetic reconnection \citep{2002ApJ...571L..63R}. 

Using high-resolution imaging observations, the standard and erupting (or blowout) jet dichotomy \citep{2010ApJ...720..757M} has now started to be addressed in $^3$He-rich SEP events \citep{2015ICRC...34...49K,2018ApJ...852...76B,2020ApJS..246...42W}. It is thought that blowout jets involve more complex reconnection processes than standard, straight jets. A twisted flux rope at the jet source, carried by a mini-filament, has been considered as a triggering mechanism for blowout jets. The events with high enrichments in both $^3$He and heavy ions including Fe are associated with blowout jets with helical structure and cool mini filaments at the base of the jets \citep{2016ApJ...823..138M,2016AN....337.1024I,2018ApJ...852...76B}. The jets in these events have shown an unwinding motion. Figure \ref{fig:3} presents an example of helical jets in three $^3$He-rich SEP events that show high $^3$He and Fe enrichments. The unwinding motions in helical jets have been related to the generation/propagation of (torsional) Alfv{\'e}n waves \citep[e.g.,][]{2011ApJ...728..103L,2015ApJ...798L..10L} which in turn can play a role in ion acceleration (see Section \ref{sec:8}).

\begin{figure*}[t!]
\centering
$\begin{array}{rl}
    \includegraphics[trim=0 80 260
105,clip=true,width=0.59\textwidth]{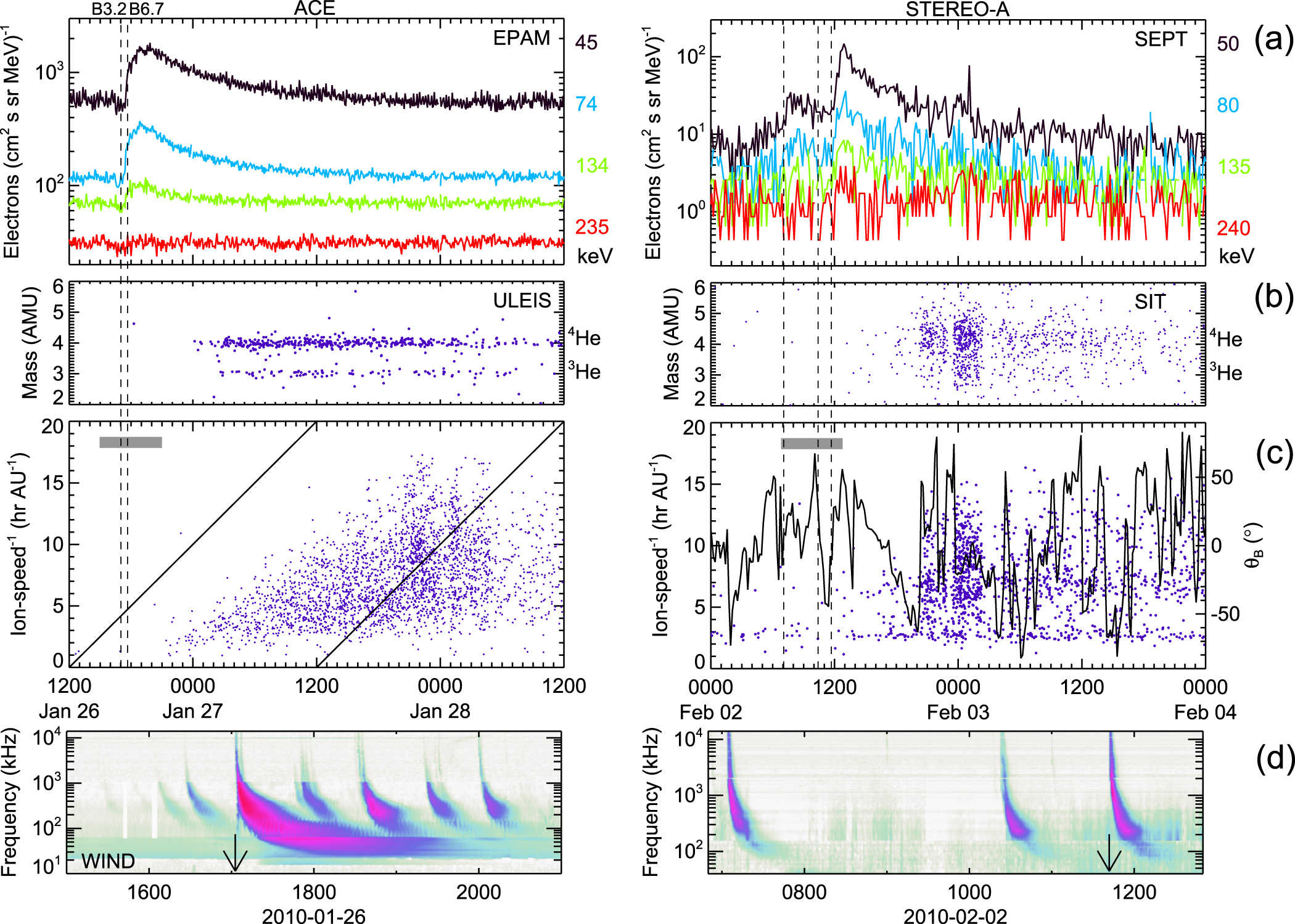} &
    \includegraphics[trim=210 0 98
0,clip=true,width=0.4\textwidth]{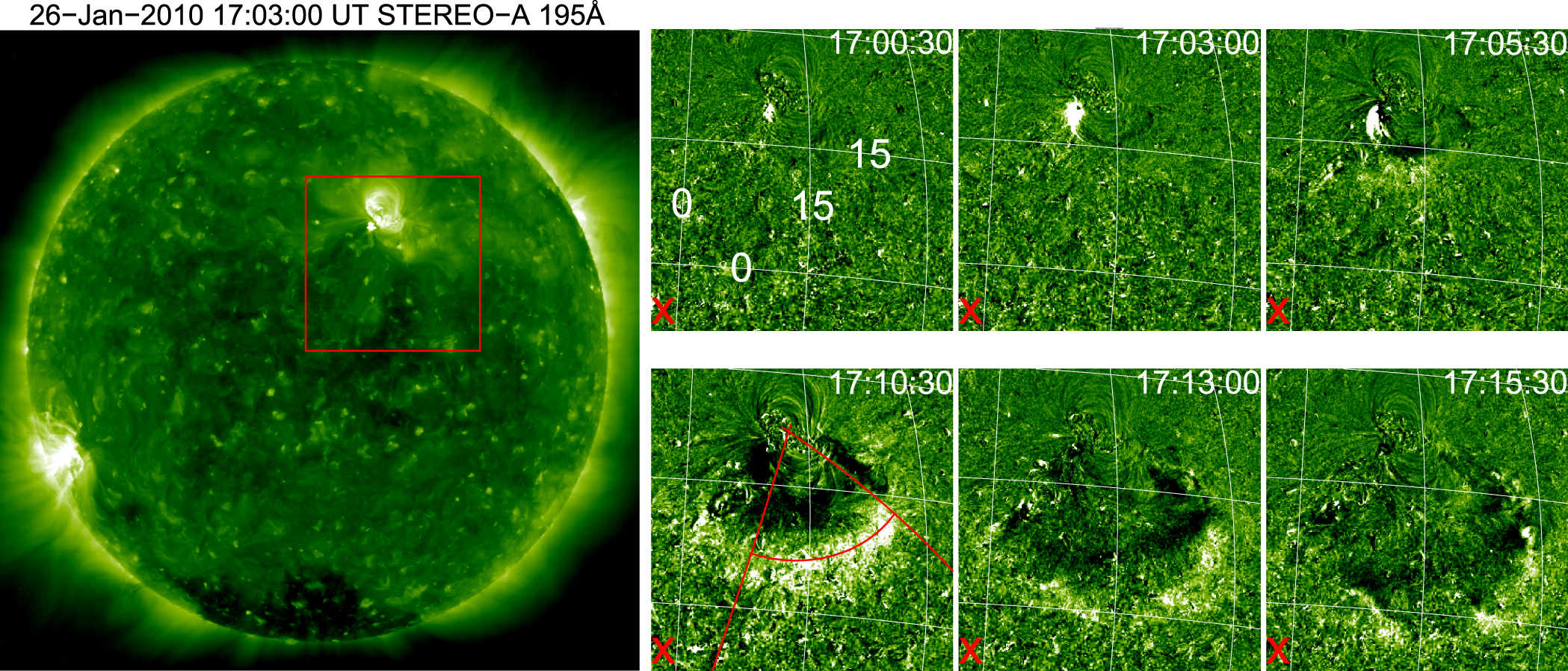}
\end{array}$
\caption{(Left) The 2010 January 26 $^3$He-rich SEP event measured on ACE. Blue dots represent 0.4--10\,MeV\,nucleon$^{-1}$ ions in the mass range 2--6\,amu (upper panel) and 0.03--10\,MeV\,nucleon$^{-1}$ ions in the mass range 10--70\,amu (lower panel). Dashed vertical lines mark GOES X-ray B3.2 and B6.7 flares. (Right) STEREO-A 195\,{\AA} running difference images around the source AR 11042 on 2010 January 26. Red crosses indicate magnetic foot-points of ACE. During the event, ACE and STEREO-A were angularly separated by 65$^{\circ}$. Adapted from \citet{2015ApJ...812...53B}.}
\label{fig:4} 
\end{figure*}

\section{EUV waves} 
\label{sec:4}
Surprisingly, jets in many $^3$He-rich events are accompanied by large-scale propagating EUV waves \citep{2015ApJ...812...53B,2016ApJ...833...63B,2015ApJ...806..235N}. Figure \ref{fig:4} shows the 2010 January 26 $^3$He-rich SEP event with the EUV wave that started as a jet. Though the nature of the EUV waves has been controversial, presently it is believed that they are true magnetosonic waves \citep[see][for a review]{2015LRSP...12....3W}. If EUV waves steepen to shocks, they may affect the energetic ion production in $^3$He-rich events \citep{2016JPhCS.767a2002B}. However, half of the EUV waves in these events are accompanied by slow CMEs ($\lesssim$300\,km\,s$^{-1}$). \citet{2019SoPh..294...37R,2019SoPh..294..141R} have suggested that events associated with fast narrow CMEs may involve CME-driven shock reacceleration of ions from magnetic reconnection. There is a tendency for $^3$He-rich events with jets to have a rounded $^3$He and Fe spectra towards low energies and for events with coronal waves to have a power-law spectra \citep{2015ApJ...806..235N,2016JPhCS.767a2002B}. It has been suggested that rounded spectra arise from a primary mechanism of $^3$He (Fe) enrichment and power laws involve a further stage of acceleration \citep{2000ApJ...545L.157M,2002ApJ...574.1039M}. It is presently unclear whether broad longitudinal distribution \citep{2013ApJ...762...54W,2015ApJ...806..235N,2016ApJ...833...63B,2017ApJ...846..107Z} or delayed ion injection \citep{2016A&A...585A.119W} measured in several $^3$He-rich events may be related to EUV waves. We note that all wide-spread $^3$He-rich events reported by \citet{2013ApJ...762...54W} were found to be accompanied by coronal waves in their source flare \citep{2016ApJ...833...63B,2015ApJ...806..235N}.

%or the waves may be responsible for broad longitudinal distributions measured in several $^3$He-rich events \citep{2013ApJ...762...54W,2015ApJ...806..235N,2016ApJ...833...63B,2017ApJ...846..107Z}. 

\begin{figure*}[t!]
\centering
% Use the relevant command to insert your figure file.
% For example, with the graphicx package use
\includegraphics[width=0.96\textwidth]{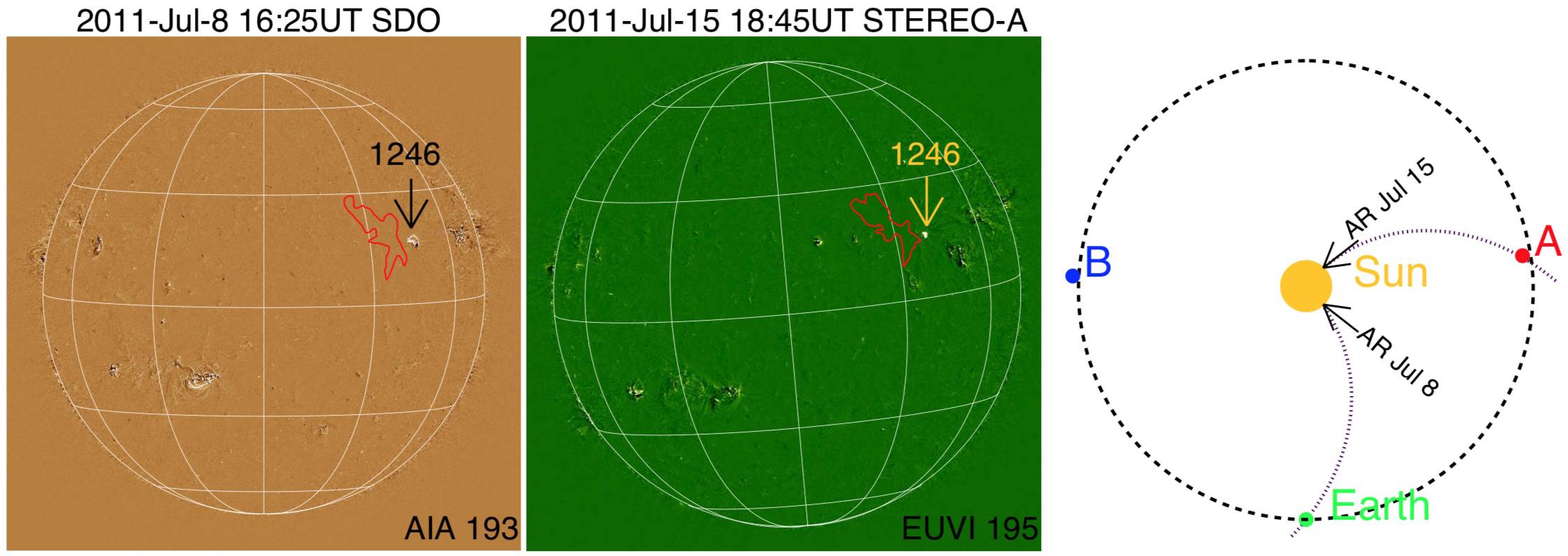}\\ \vspace{2 mm}
\includegraphics[trim=0 0 0
23.5,clip=true, width=0.496\textwidth]{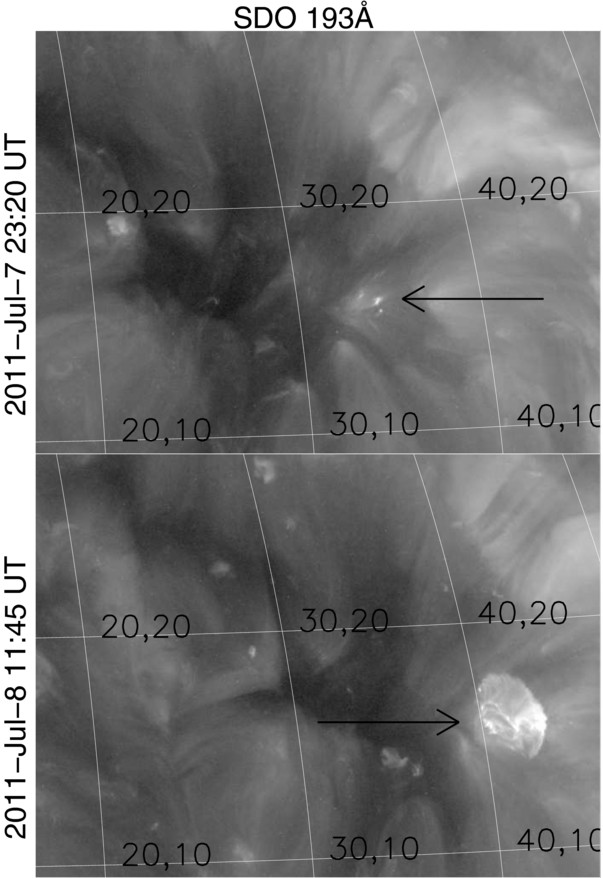}
\includegraphics[trim=0 18.5 0
0,clip=true,width=0.496\textwidth]{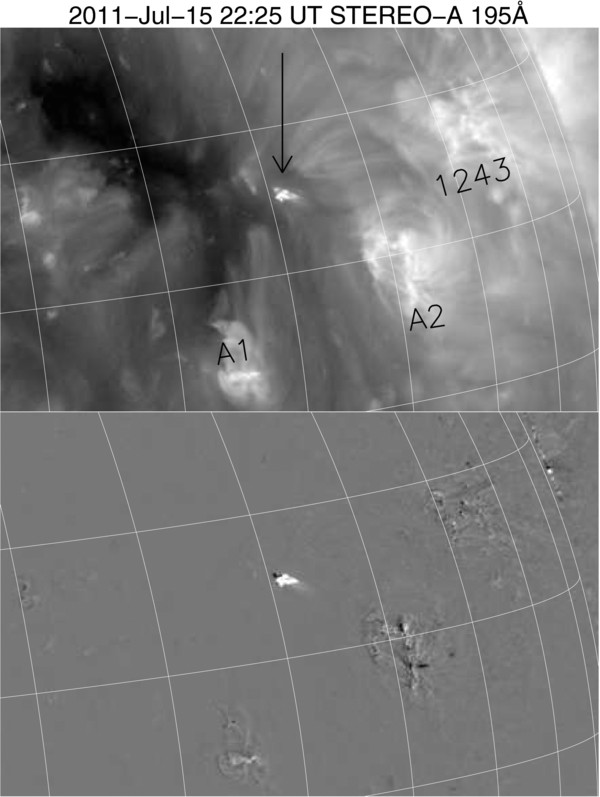}
% figure caption is below the figure
\caption{(Top) EUV running difference images of the solar disk (left, middle); the location of the spacecraft and the source AR on July 8 and July 15 (right). Arrows point to the long-lived $^3$He-rich SEP source AR11246 at the boundary of the coronal hole (red contour). The $^3$He-rich SEP events from AR11246 were measured on 2011 July 9 by ACE and on 2011 July 16 by STEREO-A. (Bottom) EUV images of a flare/jet from AR11246 in the 2011 July 9 and 2011 July 16 events. Adapted from \citet{2014ApJ...786...71B,2015JPhCS.642a2002B}. }
\label{fig:5}       % Give a unique label
\end{figure*}

\section{Long-lived sources}
\label{sec:5}
Measurements with single spacecraft have shown recurrent $^3$He-rich SEP events associated with the same source region over relatively short time ($\sim$1--2 day) periods \citep{1986ApJ...308..902R,1996GeoRL..23.1219M,1999ApJ...525L.133M,2000ApJ...545L.157M,2006ApJ...639..495W,2006ApJ...648.1247P,2015A&A...580A..16C} presumably due to loss of the magnetic connection to the flare site. Multipoint observations with angularly separated spacecraft have shown that a solar source may produce $^3$He-rich SEP events at least for a quarter of a solar rotation \citep[see Figure \ref{fig:5};][]{2013arXiv1307.6342B,2014ApJ...786...71B,2015JPhCS.642a2002B}. This indicates that more persistent conditions for ion acceleration may exist in $^3$He-rich SEP sources \citep{2006ApJ...639..495W,2006ApJ...648.1247P,2014ApJ...786...71B}. Note, multi-day periods of the nearly continuous presence of energetic $^3$He have been reported around the time of solar maximum often with no individual $^3$He-rich events resolved \citep{2007SSRv..130..231M}. Such long periods have been interpreted as a combination of continuous emission of energetic $^3$He ions and their confinement in interplanetary magnetic field structures \citep{2008ApJS..176..497K}.

\section{Photospheric source}
\label{sec:6}
\begin{figure*}[t!]
\centering
$\begin{array}{rl}
    \includegraphics[trim=4 5 325
158,clip=true,width=0.265\textwidth]{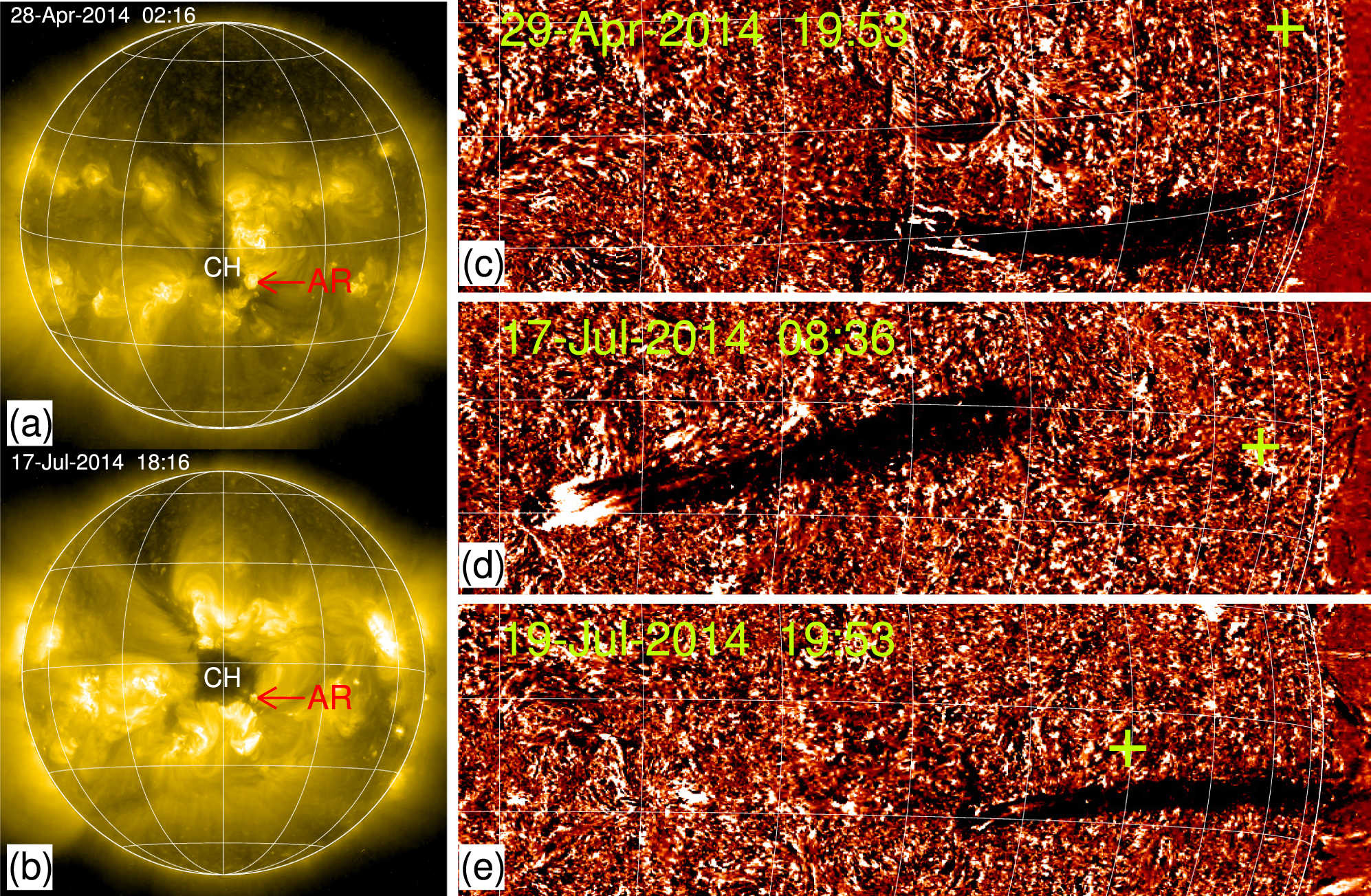} &
    \includegraphics[trim=177 106 34
104,clip=true,width=0.70\textwidth]{apjaa9d8ff7_hr.jpg}
\end{array}$
\caption{(Left) STEREO-A 284\,{\AA} EUV image of the source AR on the border of a coronal hole for the STEREO-A 2014 July 17 $^3$He-rich SEP event. It appears that the coronal hole is an extension of a polar coronal hole through a narrow corridor. (Right) STEREO-A 304\,{\AA} EUV running difference image showing a jet with significant non-radial expansion toward the STEREO-A magnetic foot-point (green plus). Adapted from \citet{2018ApJ...852...76B}.}
\label{fig:6} 
\end{figure*}

The underlying photospheric magnetic field in $^3$He-rich SEP sources is generally unknown and has been rarely addressed \citep{2016AN....337.1024I}. In their pioneering work, \citet{1987SoPh..107..385K} have reported the event associated flares close to sunspots, in old spotless regions or bright plage regions with no flares. In further inestigations, several $^3$He-rich events were found with jets from small active regions at (near-equatorial) coronal hole boundaries \citep{2006ApJ...639..495W,2013arXiv1307.6342B,2014ApJ...786...71B,2015JPhCS.642a2002B,2018ApJ...852...76B} where magnetic field lines open to the heliosphere. Indeed, $^3$He-rich SEP events tend to be detected just before the fast wind \citep{1978ApJ...225..281Z,1996GeoRL..23.1219M,2008ApJS..176..497K,2013AIPC.1539..139B} indicating the source active regions are west of coronal holes \citep{1996GeoRL..23.1219M}. Figure \ref{fig:6} shows a jet with significant longitudinal extension from a small, compact active region at the coronal hole boundary for the 2014 July 17 $^3$He-rich SEP event. A plage region \citep{2015A&A...580A..16C,2015JPhCS.642a2002B} from dispersed sunspots or newly emerged active regions can also be the source of $^3$He-rich jets. So far only one event has been reported arising directly from a sunspot jet \citep{2008ApJ...675L.125N}. %The authors have reported a $^3$He-rich SEP event with an X-ray jet originating at the edge of the sunspot umbra. 

\begin{figure*}[t!]
\centering
% Use the relevant command to insert your figure file.
% For example, with the graphicx package use
\includegraphics[trim=18 197 255
125,clip=true,width=0.6\textwidth]{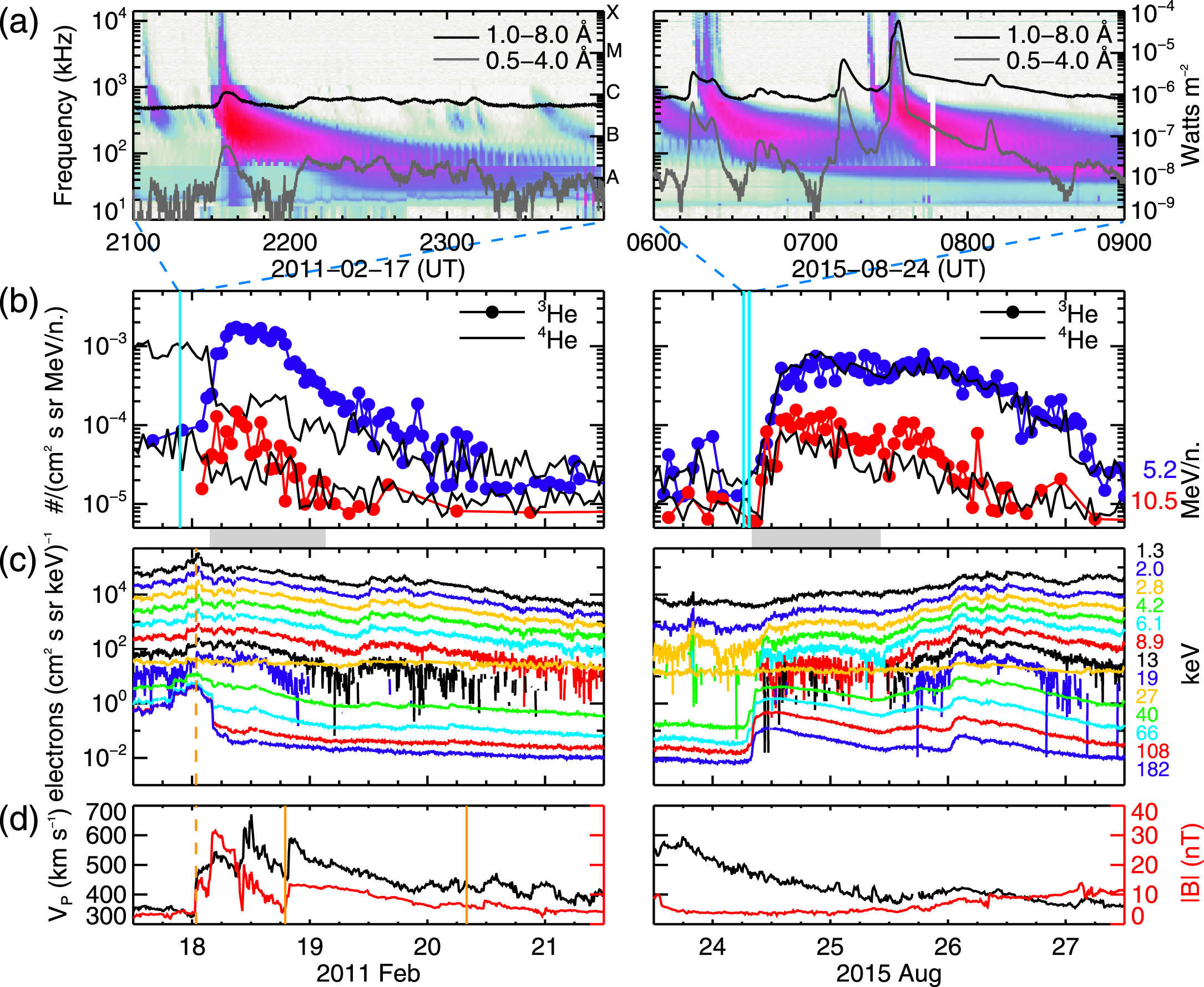}
\includegraphics[trim=18 0 255
405,clip=true,width=0.6\textwidth]{apjlaaf37ff1_hr.jpg}\\
\includegraphics[trim=0 9 340
162,clip=true,width=0.605\textwidth]{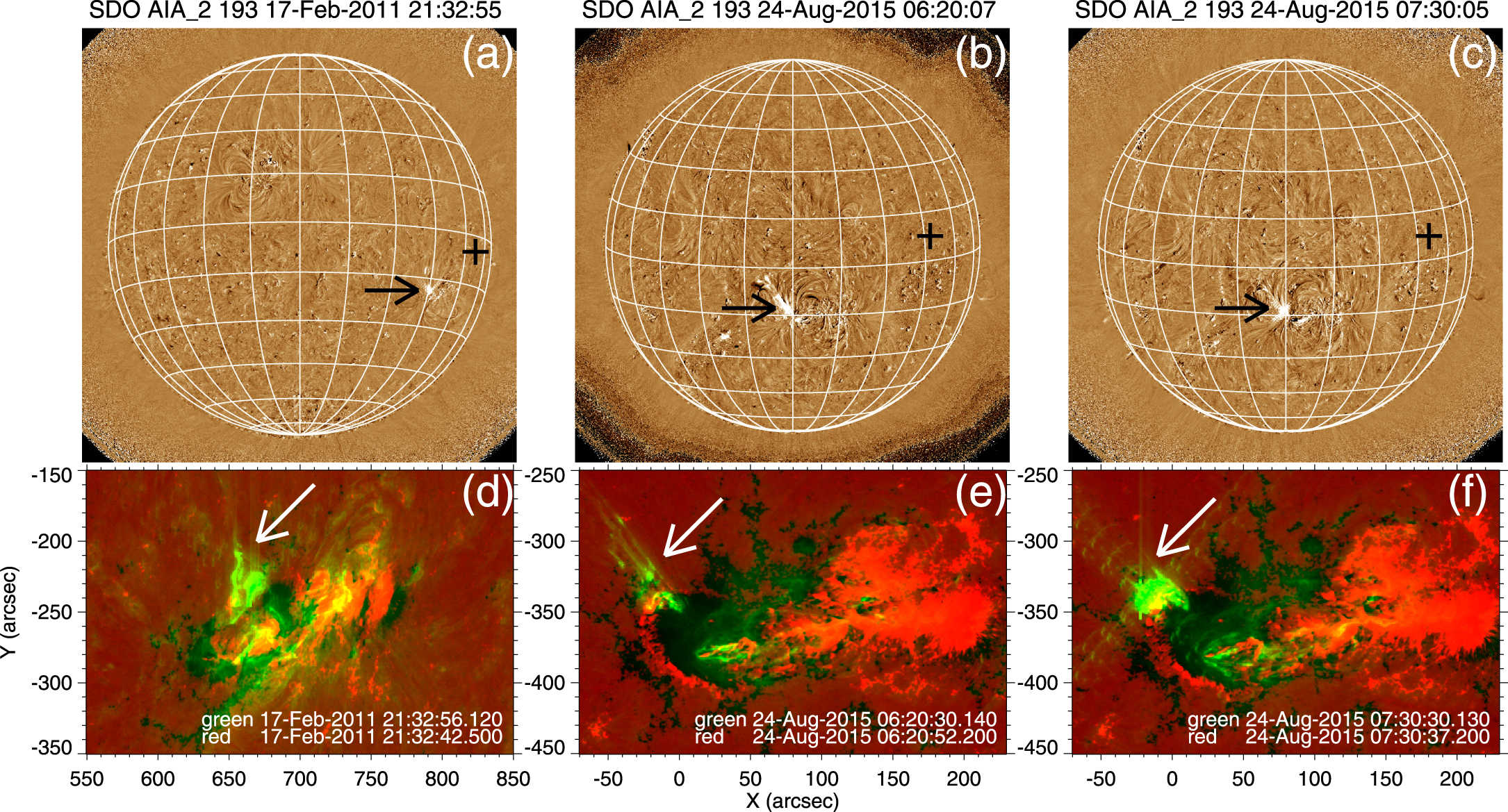}
% figure caption is below the figure
\caption{(Top) The 2011 February 18 $^3$He-rich SEP event measured on ACE at 5.2\,MeV\,nucleon$^{-1}$ (blue) and 10.5\,MeV\,nucleon$^{-1}$ (red). The solid vertical line marks the event associated C1.1 X-ray flare. (Bottom) Two-color composite image of the solar source for the 2011 February 18 event - helical jet from the boundary of the most complex and large sunspot in the solar cycle 24. Green corresponds to SDO/AIA 304\,{\AA} and red to SDO/HMI line-of-sight magnetic field. Adapted from \citet{2018ApJ...869L..21B}.}
\label{fig:7}       % Give a unique label
\end{figure*}

Recently, rare, very high energy ($>$10\,MeV\,nucleon$^{-1}$) $^3$He-rich SEP events were associated with jets from the edge of large and complex sunspots \citep{2018A&A...617A..40B,2018ApJ...869L..21B,2019ApJB}. The energy for extensive ion acceleration in these events may originate in the strong magnetic fields of large sunspots and/or from the complexity of the Sun's surface magnetic field (like shearing motions that may accumulate free energy). An example of such an event is shown in Figure \ref{fig:7}. Earlier work \citep{2002ApJ...573L..59T,2003ApJ...586.1430K,2003SoPh..214..177T} has suggested that high energy $^3$He-rich SEPs may be due to re-acceleration in coronal shocks. However, in the above-mentioned recent events no type II-radio bursts, a coronal shock signature, were observed.

\section{Temperature}
\label{sec:7}
The probable temperature of the source plasma that is accelerated in $^3$He-rich SEP events is cooler \citep[2.5--3.2\,MK;][]{2014SoPh..289.4675R,2015SoPh..290.1761R} than typical flare temperatures ($>$10\,MK). It is deduced from the measured ion abundance enhancement pattern and theoretical dependence of equilibrium ionization states on temperature \citep{1994ApJS...90..649R,2004ApJ...606..555M}. Abundances of ions such as $^{4}$He, $^{12}$C, $^{14}$N, and $^{16}$O are unenhanced in $^3$He-rich events, while heavier species like $^{20}$Ne, $^{24}$Mg, $^{28}$Si show enhancement. This suggests that C, N, O are fully ionized as $^{4}$He but Ne, Mg, Si are only partially stripped. This can happen in a temperature range of $\sim$2.5--3.2\,MK. Note that several events have been reported with enhanced $^{12}$C, $^{14}$N suggesting temperatures 0.5--1.5\,MK \citep{2002ApJ...565L..51M,2016ApJ...823..138M}. These temperatures may indicate that ions are accelerated very early \citep{2016ApJ...823..138M} and/or they are accelerated on open field lines where heating is minimal \citep{2015SoPh..290.1761R}. More direct determination of the source temperature in $^3$He-rich SEP events can be obtained from measurements of Fe charge states at low energies \citep[$<$100\,keV\,nucleon$^{-1}$;][]{2000A&A...357..716K,2001A&A...375.1075K}. \citet{2008ApJ...687..623D} have obtained temperatures in the range of 1.3 to 3\,MK using low-energy Fe measurements in several $^3$He-rich SEP events.

\begin{figure*}[t!]
\centering
% Use the relevant command to insert your figure file.
% For example, with the graphicx package use
\includegraphics[trim=0 240 0
0,clip=true,width=0.95\textwidth]{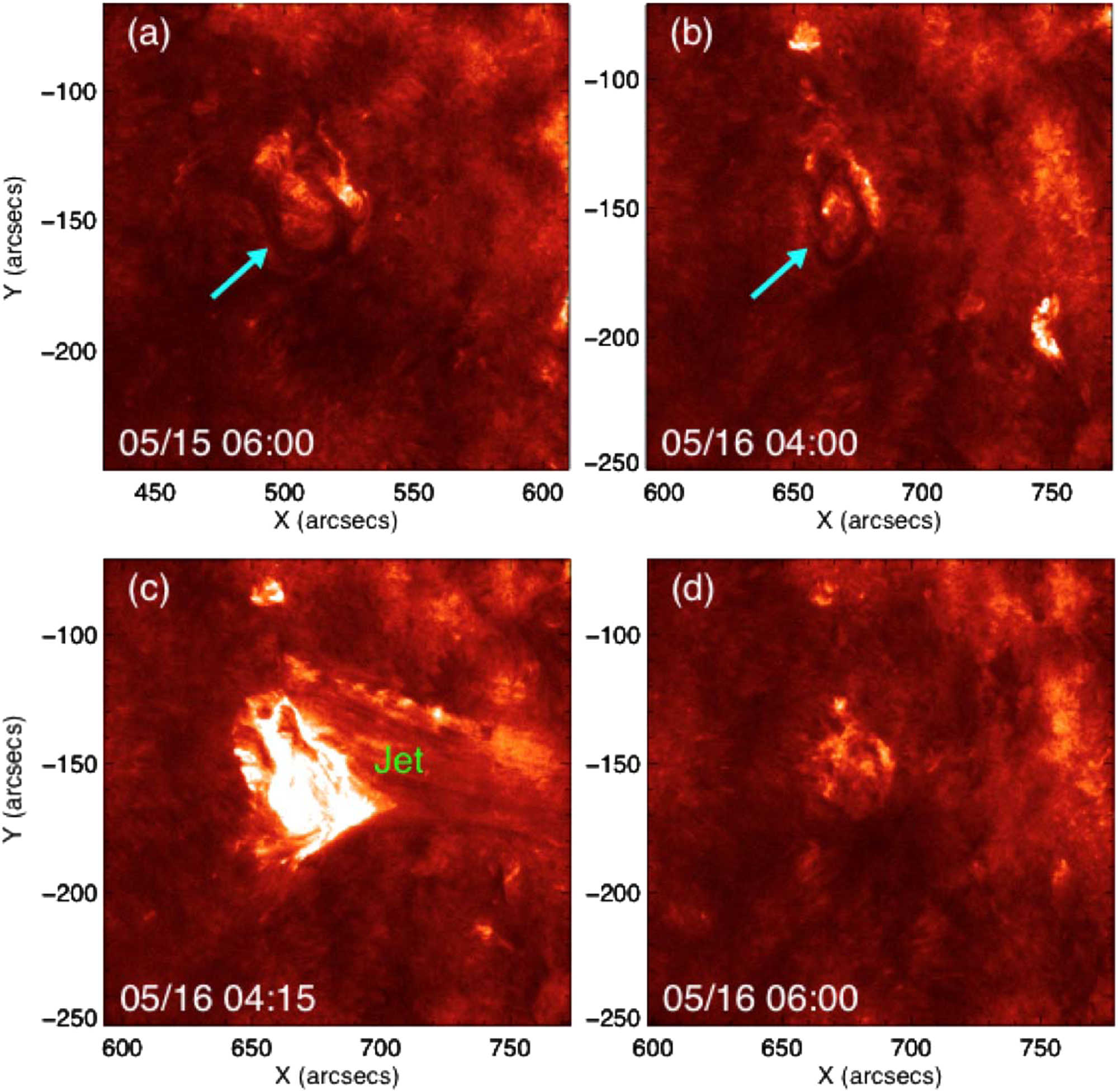}
% figure caption is below the figure
\caption{SDO/AIA 304\,{\AA} images of the solar region for the 2014 May 16 $^3$He-rich SEP event. The blue arrow shows a ring of cooler material 10\,hr  (panel a) and 15 minutes (panel b) before the jet. Adapted from \citet{2016ApJ...823..138M}.}
\label{fig:8}       % Give a unique label
\end{figure*}

The presence of cool material in the form of a mini-filament (see Figure \ref{fig:8}) may be required for acceleration of lighter species (C, N), as has been recently reported in the sources of $^3$He-rich SEP events with large $^3$He and Fe enrichments \citep{2016ApJ...823..138M,2016AN....337.1024I,2018ApJ...852...76B}. Fairly low temperatures (0.1--0.6\,MK) are required to produce observed heavy-ion enhancements in a new mechanism based on the charge per mass (Q/A) dependence of Coulomb energy losses below the Bragg peak \citep{2018ApJ...862....7M}. However, in a more complete treatment, combining Q/A dependent Coulomb losses with Q/A dependent acceleration, a similar Q/A dependence of heavy-ion abundance enhancements can be obtained, with temperatures of $\sim$1\,MK \citep{2008ApJ...681.1653K,2020ApJ...888...48K}. The source temperature $\sim$2\,MK determined with the Differential Emission Measure method using SDO EUV observations has been reported for one $^3$He-rich SEP event \citep{2018IAUS..335...14C}. In early studies, an abundance enhancement increase with soft X-ray temperature (in the range 10--16\,MK) has been shown \citep{1988ApJ...325L..53R} for some heavy elements (C, Mg, Si, S, Fe). The origin of the reported dependencies remains unclear if many of these species are fully stripped \citep{1988ApJ...325L..53R}. Why SEP abundances should correlate with the X-ray temperature of closed heated regions near the source? Limited statistics and possible wrongly associated events do not seem to explain the fairly high correlations. Could wave energy from hot regions affect nearby SEPs? \citet{1990ApJS...73..235R} has suggested that plasma waves escaping from a closed X-ray region \citep{1983ApJ...273L..95S} accelerate ions in impulsive SEP events.The $^3$He/$^4$He ratio showed no correlation with temperature \citep{1988ApJ...327..998R}.

%\shortcites{2017NatPh..13..973K}

\section{Acceleration}
\label{sec:8}
Various processes have been proposed to explain the anomalous abundances of $^3$He-rich SEPs \citep[see review by][]{1998SSRv...86...79M}. These processes address $^3$He and heavy-ion acceleration separately. They, however, must fit together in some way, even though the magnitude of these enhancements is uncorrelated. Most models involve ion-cyclotron resonance with plasma waves. The models of cascading Alfv{\'e}n waves may account for heavy-ion acceleration \citep{1998SSRv...86...79M,2004ApJ...617L..77Z} where ions with low Q/A are accelerated with a faster rate \citep{2014ApJ...794....6E,2017ApJ...835..295K}. This is qualitatively compatible with measured enhancements of heavy ions in $^3$He-rich events \citep{2004ApJ...606..555M,2004ApJ...610..510R}. The models assume that waves generated at long wavelengths during the reconnection by relaxation of twisted non-potential magnetic fields, cascading toward shorter length scales, resonate with ions of increasing gyro-frequency or Q/A \citep{1998SSRv...86...79M}. The models of $^3$He acceleration involve plasma waves generated around the $^3$He cyclotron frequency \citep{1978ApJ...224.1048F,1992ApJ...391L.105T,2006ApJ...636..462L}. These waves are assumed to be generated by an electron current, energetic electron beams, or via coupling with low-frequency Alfv{\'e}n waves. It is remarkable that the efficient acceleration of $^3$He by ion-cyclotron resonance has been recently measured in nuclear fusion devices \citep{2017NatPh..13..973K}. 

Besides models based on cyclotron resonance, some other (non-wave) mechanisms have been proposed. These include ion fractionation by reconnection outflows followed by Fermi acceleration on multiple magnetic islands \citep{2009ApJ...700L..16D,2019AAK} and fractionation by magnetic helicity-driven DC electric fields \citep{1995ApJ...452..451H,2013MNRAS.429.2515F} followed by cyclotron resonance. In the former model, a production rate is scaled by Q/A as a power-law (Drake et al. 2009) that is consistent with measurements. The magnetic islands have been observed in the reconnection sites through direct imaging and a temperature analysis \citep{2016NatPh..12..847L}. The model with helicity-driven DC electric field \citep{2013MNRAS.429.2515F} predicts common enrichment of $^3$He and Fe, reported in events with helical jets \citep{2016ApJ...823..138M,2018ApJ...852...76B}. 

\begin{figure*}[t!]
\centering
% Use the relevant command to insert your figure file.
% For example, with the graphicx package use
\includegraphics[width=0.6\textwidth]{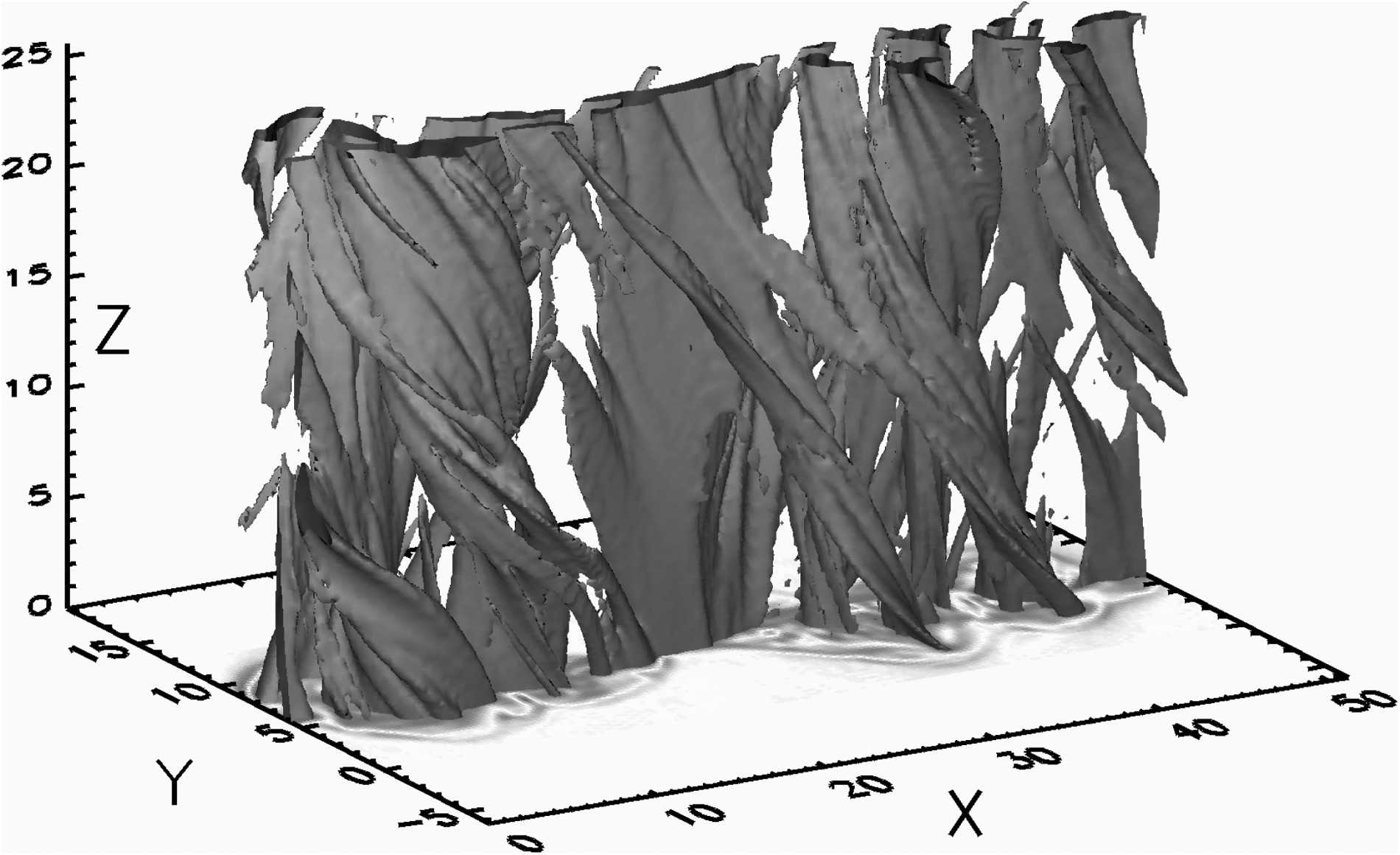}
% figure caption is below the figure
\caption{Isosurface of the electron current density that exhibit filamentary structure in the 3D simulation (from \citet{2015PhPl...22j0704D}).}
\label{fig:9}       % Give a unique label
\end{figure*}

Recent 3D simulations have shown that reconnection in the corona is intrinsically turbulent \citep{2011NatPh...7..539D}. The resulting turbulent magnetic structures, that represent magnetic islands in 2D, enhance electron \citep{2015PhPl...22j0704D} as well as ion (J. F. Drake, private communication) acceleration, facilitating particle transport. These stochastic magnetic structures are illustrated in Figure \ref{fig:9}. The observed helical morphology in $^3$He-rich jets may be a signature of these turbulent structures or torsional Alfv{\'e}n waves on large scales.

Earlier work speculated that abundance variations in source material may affect the anomalous composition of $^3$He-rich SEPs \citep{1986ApJ...303..849M,1990ApJ...357..259R}. \citet{1989ApJ...343..511W} has suggested that variations in the source material can be discernable with EUV spectral line observations. Small field-of-view of EUV spectrographs may be a limiting factor for systematic investigation of EUV lines in $^3$He-rich SEP sources, but the intentional pointing to the magnetically connected flares may be helpful \citep{2018A&A...617A..40B}. For example, proposals can be submitted for coordinated observations of the EUV Imaging Spectrometer (EIS) on Hinode and Interface Region Imaging Spectrograph (IRIS) to investigate jets in $^3$He-rich SEP sources (https://hinode.msfc.nasa.gov/hops.html).
 
\section{Summary}
\label{sec:9}
Discovered more than 50 years ago, $^3$He-rich SEP events are still poorly understood. It is mainly because of their low intensities, short duration, association with minor flares, and the requirement for a relatively accurate magnetic connection to a small size source on the Sun. Furthermore, energetic ions in $^3$He-rich events do not show flare signatures as energetic electrons through the hard X-ray bremsstrahlung or radio emissions. Signatures of ion acceleration in EUV spectral lines in large X-ray flares have been recently investigated \citep[e.g.,][]{2016A&A...590A..99J} and may find application for sources of $^3$He-rich SEPs.

We expect that new missions at a close distance of the Sun, Parker Solar Probe (launched in 2018) and Solar Orbiter (launched in 2020) will revolutionize our view on processes responsible for ion acceleration in $^3$He-rich SEP events. For example, observations from an unprecedentedly close distance ($\sim$10--35 solar radii; compare 215 solar radii from 1\,au) to the Sun will remove uncertainties due to interplanetary propagation effects and magnetic connection. Such observations may reveal new solar sources of $^3$He-rich SEPs. One such source may be a local acceleration of $^3$He-rich SEPs in the solar wind (J. F. Drake, private communication). 

\begin{acknowledgements}
Figs. \ref{fig:1}--\ref{fig:4}, \ref{fig:6}--\ref{fig:8} reproduced with the permission of AAS. Fig. \ref{fig:9} reprinted with the permission of AIP Publishing. The author thanks Davina Innes and Glenn Mason for the careful reading of the manuscript. The paper also benefit greatly from the discussions at the ISSI International Team ID 425 'Origins of $^3$He-rich SEPs' led by R. Bu\v{c}\'ik and J. F. Drake. The work was partly supported by Deutsche Forschungsgemeinschaft (DFG) grant BU 3115/4-1. 
%If you'd like to thank anyone, place your comments here
%and remove the percent signs.
\end{acknowledgements}

% Authors must disclose all relationships or interests that 
% could have direct or potential influence or impart bias on 
% the work: 
%
% \section*{Conflict of interest}
%
% The authors declare that they have no conflict of interest.

% BibTeX users please use one of
%\bibliographystyle{spbasic}      % basic style, author-year citations
%\bibliographystyle{spmpsci}      % mathematics and physical sciences
%\bibliographystyle{spphys}      % APS-like style for physics
%\bibliographystyle{plainnat} 
\bibliographystyle{abbrvnat}
\bibliography{ssrv}   % name your BibTeX data base

% Non-BibTeX users please use
%\begin{thebibliography}{}
%
% and use \bibitem to create references. Consult the Instructions
% for authors for reference list style.
%
%\bibitem{RefJ}
% Format for Journal Reference
%Author, Article title, Journal, Volume, page numbers (year)
% Format for books
%\bibitem{RefB}
%Author, Book title, page numbers. Publisher, place (year)
% etc
%\end{thebibliography}

\end{document}